# Batch-FPM: Random batch-update multi-parameter physical Fourier ptychography neural network

**Ruiqing Sun[a], Delong Yang[a], Yiyan Su[c], Shaohui Zhang[a, *], and Qun Hao[a, b, *]**

[a]School of Optics and Photonics, Beijing Institute of Technology, Beijing 100081, China
[b]Changchun University of Science and Technology, Changchun 130022, China
[c]Department of Health Statistics, School of Public Health, Shanxi MedicalUniversity, 56 South XinJian Road, Taiyuan 030001, Shanxi Province, China

**Abstract:** Fourier Ptychographic Microscopy (FPM) is a computational imaging technique that enables high-resolution imaging over a large field of view. However, its application in the biomedical field has been limited due to the long image reconstruction time and poor noise robustness. In this paper, we propose a fast and robust FPM reconstruction method based on physical neural networks with batch update stochastic gradient descent (SGD) optimization strategy, capable of achieving attractive results with low single-to-noise ratio and correcting multiple system parameters simultaneously. Our method leverages a random batch optimization approach, breaks away from the fixed sequential iterative order and gives greater attention to high-frequency information. The proposed method has better convergence performance even for low signal-to-noise ratio data sets, such as low exposure time dark-field images. As a result, it can greatly increase the image recording and result reconstruction speed without any additional hardware modifications. By utilizing advanced deep learning optimizers and perform parallel computational scheme, our method enhances GPU computational efficiency, significantly reducing reconstruction costs. Experimental results demonstrate that our method achieves near real-time digital refocusing of a 1024 × 1024 pixels region of interest on consumer-grade GPUs. This approach significantly improves temporal resolution (by reducing the exposure time of dark-field images), noise resistance, and reconstruction speed, and therefore can efficiently promote the practical application of FPM in clinical diagnostics, digital pathology, and biomedical research, etc. In addition, we believe our algorithm scheme can help researchers quickly validate and implement FPM-related ideas. We invite requests for the full code via email.

**Keywords**: Physics-based neural network, computational imaging, Fourier Ptychographic microscopy.

**\*** Shaohui Zhang**,** E-mail: zhangshaohui@bit.edu.cn
**\*** Qun Hao, E-mail: qhao@bit.edu.cn

## 1. Introduction

High spatial bandwidth product (SBP) imaging is essential in pathology for examining tissue sections to observe cell morphology, size, arrangement, and nuclei, which is crucial for clinical diagnoses [1-5]. Fourier Ptychography Microscopy (FPM) is a representative scheme achieving large SBP imaging through computational imaging methods rather than mechanical scanning [6]. It realizes the improvement of imaging space-bandwidth product by collecting a series of intensity images at different illumination angles provided by the LED board and performing phase recovery and aperture synthesis in the frequency domain. Since FPM uses a low NA objective lens with relatively large imaging field of view, it does not or seldom need mechanical scanning to expand the imaging field of view. The inverse problem in FPM, consisting of phase retrieval and aperture synthesis, is a complex, nonlinear process including multiple cycles alternating projections between real and Fourier space and corresponding constraints to accelerate the convergence speed. However, unlike the fast spatial domain images stitching, the Fourier domain phase retrieval and stitching in FPM requires multiple optimization iteration loops to achieve, which is usually very time-consuming, and sensitive to data noise. It will make the overall time of FPM imaging and reconstructing the final image very long, as well as the uncertainty of reconstruction result. This is also one of the vital reasons why FPM has been developed very well in the academic field for more than ten years since its advent, but there is few engineering products used in real pathological analysis scenarios.

To accelerate the overall imaging speed of FPM, some studies have deployed the entire computational process onto GPUs, significantly speeding up the reconstruction process [9,10]. Other approaches have improved image acquisition efficiency by increasing illumination power or using multiple additional eyepieces [7,28]. However, these methods undoubtedly increase the cost of FPM, reducing its competitiveness in the market. Despite these advancements, FPM still requires tens of seconds or even minutes for image acquisition and reconstruction to achieve high-quality imaging results, which is unacceptable for some biomedical tasks that demand timely observations [8].

As previously mentioned, time consumption is not the only factor limiting the clinical application of FPM-related technologies. Due to the non-convex nature of the phase retrieval problem, many reconstruction algorithms, including alternating projection methods, cannot guarantee convergence to a global optimum even under ideal conditions [35,22,23]. The ePIE algorithm [34], an alternating projection method derived from the Ptychographic Iterative Engine, is commonly used in FPM as a standard approach to recover phase information lost during data acquisition. As an incremental gradient method, ePIE iteratively improves the estimation of sample parameters using the collected data. Theoretically, ePIE can achieve satisfactory estimates at a reasonable speed and has been applied to various imaging algorithms, including FPM. Rendenburg et al. [36] compared sequential methods with batch update approaches and found significant differences in attention given to different sample regions. In frameworks like FPM, which update progressively from the center to the periphery, high-frequency information at the periphery receives significantly less attention than low-frequency information at the center. However, obtaining clear high-frequency details is often crucial.

While batch-update methods like RAAP and DP theoretically offer advantages over sequential updates, in practice, these traditional batch-update methods do not always outperform sequential updates. This discrepancy may involve issues related to the selection of initial values and optimizers. Moreover, FPM involves numerous high-angle dark-field images, which typically have a lower signal-to-noise ratio compared to bright-field images, making the reconstruction process referencing these data unstable. In such cases, both sequential and batch-update methods may struggle to deliver satisfactory reconstruction results [36,37]. Fortunately, many reconstruction algorithms have been proposed, leveraging advanced optimization techniques from fields like machine learning, including adaptive step sizes and momentum acceleration, to improve FPM's noise resistance and reconstruction speed [13,29,30]. However, these methods still do not give sufficient attention to dark-field information, resulting in the need for many iterations to ensure reconstruction quality.

In addition to the low signal-to-noise ratio in dark-field images caused by inexpensive image sensors or low exposure times, non-ideal system parameters such as LED position deviations, defocus, and pupil aberrations can also degrade or even fail the reconstruction results. More importantly, since these system parameters are always interrelated during inverse optimization, it is difficult to effectively separate and accurately recover them using analytical gradient expressions in traditional optimization frameworks. To address this, many studies have used deep learning frameworks to model the original imaging process [19,31,32]. By utilizing automatic differentiation and advanced optimizers in the deep learning field, joint correction of multiple system error parameters can be achieved. However, these methods still follow traditional optimization algorithms with fixed iteration sequences, which often leads to error accumulation during iterations, ultimately compromising overall imaging quality.

In this paper, we propose a batch-update multi-parameter physical Fourier ptychography neural network that enhances the practical value of FPM by improving both reconstruction speed and image quality, as shown in Fig. 1. Inherited from physical neural networks, our method can simultaneously correct multiple parameters, including defocus distance, pupil function and illumination intensity. It also leverages advanced optimizers from the deep learning field to enhance its noise resistance. Unlike previous sequential update methods, we modify the model structure to be compatible with batch-based optimization methods. For parameters such as pupil function and defocus distance, which are independent of illumination angles in joint correction, using random batch updates can effectively reduce noise influence while accelerating convergence. Unlike previous sequential update methods that placed significant weight on the central part while neglecting the edges [36], our approach gives more attention to dark-field information. This focus is crucial for the accurate reconstruction of high-frequency details, which significantly enhance the overall quality of the reconstructed images. Specifically, our use of random batch updates functions as an additional attention mechanism, ensuring that the proportion of dark-field images in each randomly selected batch matches that in the entire dataset. This consistent contribution of dark-field images to each gradient calculation helps mitigate the error accumulation caused by dark-field noise. Consequently, our algorithm can reconstruct the complex amplitude of samples with fewer iterations and shorter acquisition times, all without requiring any hardware modifications. To further increase the practical value of FPM, we have also achieved more comprehensive utilization of GPU computational power, significantly reducing both the economic and time costs of the image reconstruction process. Our method enables near-real-time digital refocusing reconstruction of regions of interest with a 1024 × 1024 pixel size on relatively affordable consumer-grade GPUs. This approach provides an innovative solution for advancing the widespread adoption and application of FPM systems in clinical diagnostics and biomedical research, while also helping researchers conduct FP-related experimental and theoretical studies more quickly and efficiently.

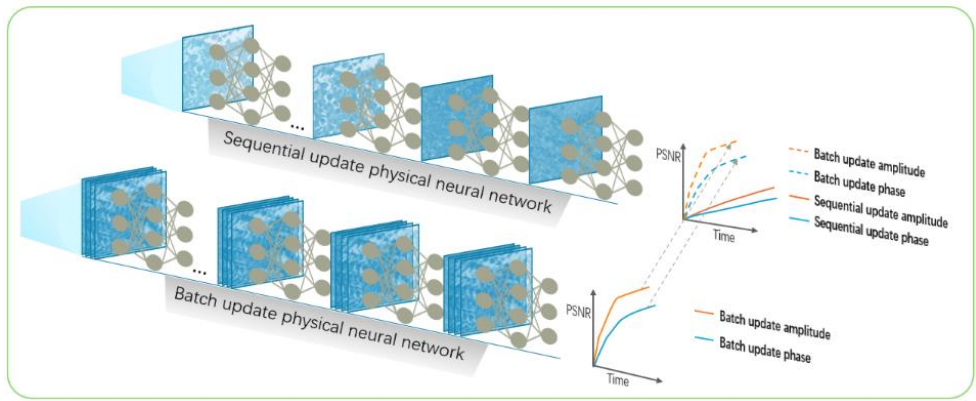

**Fig. 1. The figure shows the difference between our proposed method and previous sequential methods.**

Overall, our work achieves the following breakthroughs: 1. Achieving lower signal-to-noise ratio imaging, thereby improving the temporal resolution of FPM. 2. Implementing simultaneous multi-parameter correction based on random batch updates, enhancing the robustness of FPM imaging. 3. Increasing the reconstruction speed of FPM. This paper is organized as follows: In Section 2.1, we introduce the forward propagation and reconstruction principles of FPM. Section 2.2 presents our improved multi-parameter physical neural network. In Section 2.3, we analyze errors and optimization directions in the iterative process and provide the theoretical basis for our method. Section 3.1 presents simulation experimental results to quantitatively analyze the performance of the proposed method. Section 3.2 shows actual experimental results using the USFA resolution target as an observation object, which are cross validated with the simulation results. Section 3.3 displays our results of the pathology slice and digital refocusing of different regions of interest, demonstrating the practical value of our method. Discussion and conclusion are provided in Chapter 4.

## 2. Method

*2.1 The principle of FPM reconstruction*

The classical FPM system replaces the illumination source of a conventional microscope with a programmable LED board [6]. Unlike mechanical scanning, FPM captures a series of LR images at different illumination angles by sequentially turning on LEDs positioned at various points to scan the sample spectrum. LED illuminations from different positions act as localized plane waves with varying angles of incidence. In this process, the observed sample is considered a flat two-dimensional thin layer, and the outgoing wave across the sample can be expressed by Eq. (1). Here, $o(r)$ represents the object's spectrum, and $r = (x, y)$ represents the sample level coordinates. $\varphi(r)$ represents the phase modulation distribution, and $\mu(r)$ represents the phase's absorption distribution. The series of LR images captured by the camera during the spectral scanning process can be represented by Eq. (2). Here, $I$ denotes the image intensity, $F$ denotes the two-dimensional

Fourier transform, $k_m = (\sin\theta_{xm}/\lambda, \sin\theta_{ym}/\lambda)$ represents the spectral shift, $\lambda$ denotes the wavelength of the incident light, $(\theta_{xm}, \theta_{ym})$ denotes the angle of tilt of the incident light, and $P(k)$ denotes the pupil function. In the actual observation process, it is usually necessary to manually or automatically focus the sample. Inaccurate focusing often leads to a decrease in the FPM imaging quality. Additionally, due to manufacturing limitations, the luminous intensity of LEDs in different positions often varies, affecting imaging quality. To further improve the imaging effect, Eq. (2) is rewritten as Eq. (3), where $H(k,z) = \exp(j\frac{2\pi}{\lambda} \cdot z \cdot \sqrt{1 - k_x^2 - k_y^2})$ is used to model the effect of defocus distance, and $\gamma_i$ is a real number used to represent the difference in luminous intensity.

$$o(r) = exp\,(i\varphi(r) - \mu(r)) \tag{1}$$

$$I_i = |F^{-1}(o(k - k_{mi})P(k))|^2, i = 1,2,\dots,n \tag{2}$$

$$I_i = \gamma_i |F^{-1}(o(k - k_{mi})P(k)H(k,z))|^2, i = 1,2,\dots,n \tag{3}$$

Most current methods treat the reconstruction of HR images as a nonlinear phase retrieval problem and solve it iteratively using PIE-based or global optimization methods [11-13]. Its cost function can be expressed by Eq. (4), where $\|.\|^2$ denotes the Euclidean distance, N denotes the number of captured low-resolution images, $I_{pred_i}$ denotes the predicted value of the sample spectrum, and $I_{gt_i}$ denotes the measured value captured by the camera.

$$min\,\varepsilon = \sum_{i=1}^{N} \|I_{pred_i} - I_{gt_i}\|^2 \tag{4}$$

*2.2 multi-parameter physical neural network*

To achieve high-quality imaging results, accurate physical modeling of the imaging process is crucial. Xiaoze Ou et al. proposed an embedded optical pupil function recovery method to restore the optical pupil function of the system [14]. Jiasong Sun and Eckert Regina suggested methods to correct the positional deviations of LED arrays [15-17]. While these methods yield satisfactory results for single-parameter optimization, the effects of different parameters in the imaging process are often coupled, making simultaneous optimization of multiple parameters challenging [18]. This significantly slows down the FPM reconstruction process, reducing its utility in biomedical applications. Fortunately, Delong Yang et al. [19] utilized a physical neural network to simultaneously model multiple parameters and employed numerical differentiation instead of analytical differentiation for gradient computation, enabling the concurrent optimization of multiple parameters. Additionally, physical neural networks can easily leverage advanced optimizers, such as AdamW, Nadam, and RMSprop, integrated within deep learning frameworks, and can introduce additional regularization terms as needed. To enable the network to dynamically select and update batch sizes based on the available GPU memory, we extended the original work, as shown in Figure 2. The improved network processes inputs in batches and uses complex-type tensors provided by PyTorch. From an information processing perspective, we reformulate the optimization objective of FPM as shown in Equation (5), where Loss represents the loss function during parameter updates. Thanks to numerical differentiation, the neural network's loss function can be flexibly chosen according to the specific backend task [7,20].

$$min\ Loss(I_{pred}, I_{gt}) = \sum_{i=1}^{N} Loss(I_{pred_i}, I_{gt_i}) \tag{5}$$

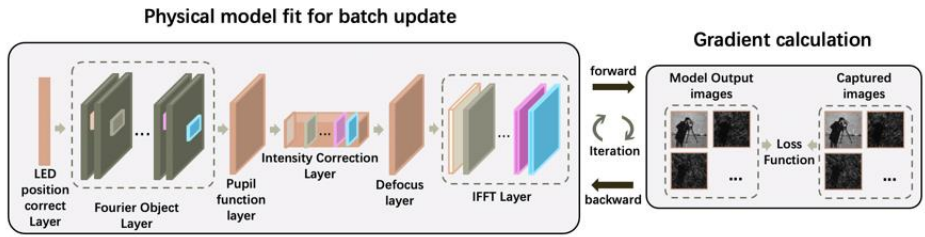

**Fig. 2. Fast and robust Fourier ptychography multi-parameter neural network with batches update.**

*2.3 multi-parameter reconstruction based on random batch-update*

Existing state-of-the-art FPM reconstruction algorithms can be broadly categorized into global gradient methods and incremental gradient methods. These methods are equally applicable to the updating process of the physical neural network. The updating process of the global gradient method each time can be described by Eq. (6), where $W$ denotes the parameter to be updated, $\nabla_w$ denotes the gradient of the parameter to be updated, and $\alpha$ is an adjustable variable to control the step size of each update. $k$ denotes the current number of update rounds, and $N$ denotes the number of low-resolution images collected. The global gradient method first sequentially calculates the losses corresponding to different measurements during the update process and updates them after summing them up [13].

$$W^{k+1} = W^k - \alpha^k \sum_{i=1}^{N} \nabla_W Loss(I_{pred_i}, I_{gt_i}) \tag{6}$$

Although global gradient methods can steer each parameter update towards the global optimum, they can result in excessively long computation times when dealing with a large number of LR images. In contrast, incremental methods, which update the parameters immediately after each gradient computation, exhibit a faster convergence rate in practice and thus have garnered more attention [6,13,19]. The updating process is shown in Eq. (7).

$$W^{k+1} = W^k - \alpha^k \nabla_W Loss(I_{pred_i}, I_{gt_i}) \ i = 1,2 \dots, N \tag{7}$$

By comparing Eq. (6) and (7), it is evident that the gradient of the incremental method during the updating process has a smaller variance. This explains why incremental methods usually have faster convergence [21]. However, while incremental methods can quickly converge to more accurate reconstruction results early on, they tend to exhibit a limit loop-like behavior during subsequent iterations, as $W^k = W^{k+\tau N}$ where $\tau \in \{1,2,3,\dots\}$. This oscillation is more pronounced when the data contains noise.

Although the exact reason for this behavior is not fully understood [22,23], we attempt to provide an analysis from the perspective of vector operations. Here, we rewrite Eq. (6) and (7) and consider a round of updates when all measured images are involved in one update. Let $\varepsilon$ represent the noise vector. Eq. (6) is rewritten as Eq. (8), and Eq. (7) is rewritten as Eq. (9) and (10). For each round of updates, the modulus of the noise for the global gradient method can be expressed by Eq. (11), while the modulus of the noise for the incremental method can be represented by Eq. (12). Clearly, when using the Euclidean distance as an indicator of the offset induced by the noise at each round, we have $\left\|\varepsilon_{global}^k\right\|^2 \leq \left\|\varepsilon_{incremental}^k\right\|^2$. This explains why the incremental gradient method is still susceptible to noise in the presence of sufficient data redundancy. Therefore, to achieve better convergence results, many previous works gradually reduce the update step size during the iteration process. Although this does not fundamentally eliminate the loop, it helps the algorithm converge to a point within the loop range through additional optimization rounds.

$$W^{k+1} = W^k - \alpha^k \sum_{i=1}^{N} \nabla_W Loss(I_{pred_i}, I_{gt_i}) - \alpha^k \sum_{i=1}^{N} \varepsilon_i^k \tag{8}$$

$$W_{i+1}^k = W_i^k - \alpha^k \nabla_W Loss(I_{pred_i}, I_{gt_i}) - \alpha^k \varepsilon_i^k \tag{9}$$

$$W^{k+1} = W_N^k \tag{10}$$

$$\left\|\varepsilon_{global}^k\right\|^2 = \left\|\sum_{i=1}^{N} \varepsilon_i^k\right\|^2 \tag{11}$$

$$\left\|\varepsilon_{incremental}^k\right\|^2 = \sum_{i=1}^{N} \left\|\varepsilon_i^k\right\|^2 \tag{12}$$

Previous global gradient methods or incremental gradient methods need to sequentially calculate the constraints brought by each LR image, thus updating the system parameters. This means that the loss and numerical discretization are computed based on one LR image before moving on to the next image, which undoubtedly reduces the degree of parallelism in the image reconstruction process. Many previous studies have shown that the update order of incremental methods is not constrained by the reconstruction algorithm itself [13,24]. It is theoretically feasible to compute the gradients corresponding to multiple LR images simultaneously in parallel [25-27]. Therefore, to further improve the reconstruction speed, we introduce the concept of batch size in the single update process and compute the gradients corresponding to multiple different low-resolution images in parallel. The parameter update process is rewritten as Eq. (13), where B stands for batch size.

$$W^{k+1} = W^k - \alpha^k \sum_{i=1}^{B} \nabla_W Loss(I_{pred_i}, I_{gt_i}) - \alpha^k \sum_{i=1}^{B} \varepsilon_i^k \tag{13}$$

Without changing the original order of gradient computation, a single batch update can be viewed as a global gradient optimization method within a certain range. To better understand this process, let's consider an example. Suppose there are a total of 9 LEDs with different angles, and the original update order is {1,2,3,4,5,6,7,8,9}. If we set the batch size to 3, the update order becomes {1,2,3}, {4,5,6}, {7,8,9}. Despite the increased computational parallelism, this approach approximates the use of global optimization within a region. This unavoidably introduces a larger gradient variance, leading to slower convergence, as demonstrated earlier. For this reason, unlike previous sequential updates, we randomly select the LR images contained in each batch. Since spectral overlap in FPM exists only between LEDs in neighboring positions, the randomized approach ensures the spectral range of each update is uniformly distributed, minimizing overlap. Therefore, it can be approximated as using the incremental gradient method in parallel on different localized regions. In addition, for parameters such as pupil and defocus distance, their values remain the same under different illumination angles. The noise in each acquisition process follows an independent and identical distribution, and its distribution mean can be regarded as a constant $\alpha$ that is only related to the system. Therefore, the system noise can be treated as a constant $\alpha$ when randomly sampling batches and can be directly corrected using subtraction. The sequential update noise is represented as a vector $\varepsilon = \sum_{i=1}^{N} \varepsilon_i$, which is related to a single acquisition process. Thus, for parameters such as pupil and defocus distance that are not related to the illumination angle in joint correction, using random batch updates can effectively reduce noise interference and allow for a larger initial iteration step to speed up convergence. Another benefit of stochastic selection is that it fundamentally breaks the loop that exists in the optimization process mentioned above. This makes it easier to escape local optima during optimization, leading to faster convergence and better reconstruction results. With existing deep learning frameworks, physical neural networks suitable for batch training can be easily deployed on consumer GPUs. Like the optimization process of many deep learning networks, the batch size is adjustable based on the computational resources available, allowing full utilization of these resources. Any randomly selected batch can be viewed as an unbiased estimate of the true gradient, so the batch size usually has no effect on the result. It is worth mentioning that physical neural networks are built based on physical rules and can accurately model the imaging process using only a small number of parameters. This allows our proposed optimization method to be updated using a large batch size with high parallelism even when performing WSI.

It is worth mentioning that the optimization process of FPM is highly dependent on the selection of initial values. In other words, reducing large oscillations at the beginning of the optimization is particularly important, making it challenging to randomly determine the update order. Traditional deep learning networks that implement end-to-end task learning typically rely on training sets containing large amounts of data and perform hundreds or thousands of rounds of iterations. In contrast, the iterative process of FPM usually consists of only a few to tens of rounds, and the local spectral regions of HR images are constrained by only a small number of captured images. Therefore, to improve the stability of the algorithm, in addition to using the bilinear interpolated center-light image as the initial value, we also ensure that each LR image is selected the same number of times during the actual deployment process. This effectively reduces the potential risk of the network parameters locally overfitting to one or more LR images in the frequency domain at the beginning of the optimization. Eq. (14) describes how to calculate the number of updates $n$ for each image, where epoch represents the total number of optimization rounds.

$$n = \frac{B * epoch}{N} \tag{14}$$

## 3. Experiments and results

*3.1 simulations*

Compared with actual experiments, the truth values of simulation experiments are easier to obtain. Therefore, we first used simulations to verify the performance of the fast parallel scheme and quantitatively analyzed the imaging performance of the algorithm using PSNR as an evaluation metric. We set the incident illumination wavelength to 470 nm, the image sensor pixel size to 2.4 μm, and a 4X objective lens with an NA of 0.13. As in the actual experiments, we placed an 11 × 11 LED array in the simulation at 97 mm from the sample, with the distance between neighboring LEDs being 5 mm. We use the images shown in Fig. 3 (a) and (b) as the true intensity and phase of the sample layer, with the ground truth containing 1024 × 1024 pixels. We simulated the light field propagation by executing the forward process in our physical neural network, thus obtaining 121 LR images of size 256 × 256 pixels. To more closely simulate actual experiments, we added Gaussian noise with different variances and a mean of 0 to each LR image, assuming the constant $\alpha$ mentioned in Section 2.3 to be 0. In real experiments, to improve the signal-to-noise ratio of dark-field images, different exposure times are usually set for images from different locations of

LED lights. Therefore, the variance of Gaussian noise in different images is set to $0.1 * \max(image_{LR})$, where max denotes the maximum value and $image_{LR}$ denotes the current LR image. The final HR phase and amplitude of $1024 \times 1024$ pixels are recovered by performing different methods. The effects of Gaussian noise with different variances are shown in Fig. 3.

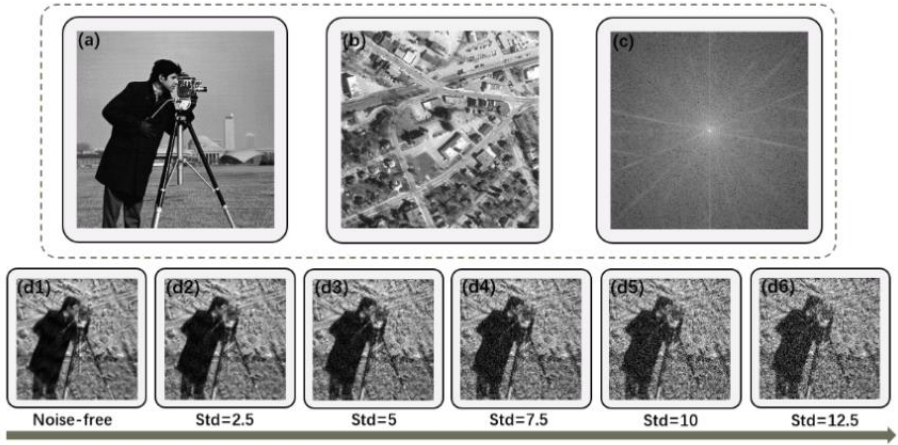

**Fig. 3. (a) The ground truth amplitude. (b) The grand truth phase. (c) The ground truth of spectral. (d*) The impact of Gaussian noise with different variances.**

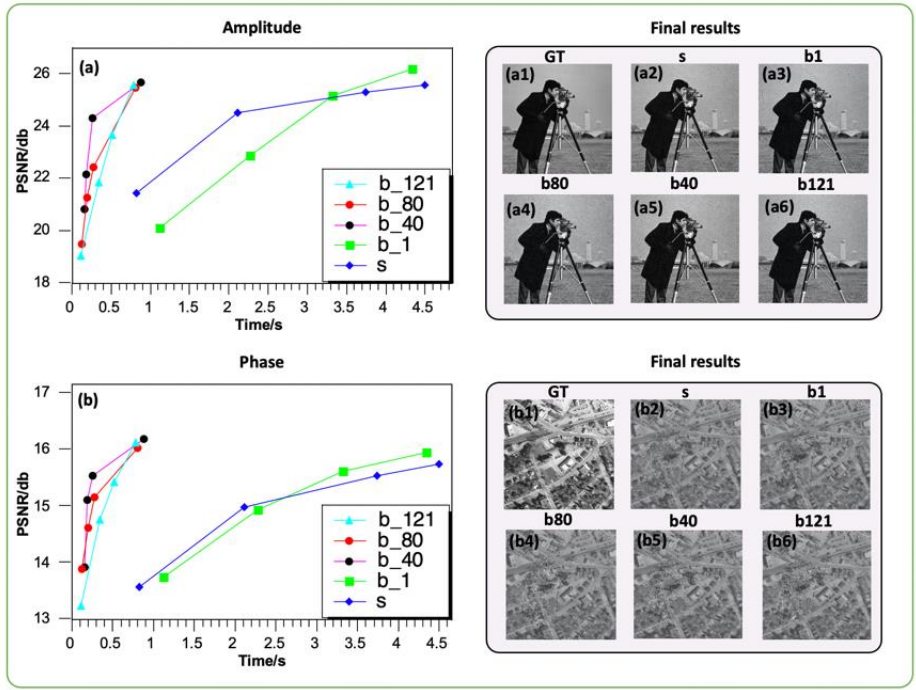

**Fig. 4. The simulation results. (a) shows the PSNR values for amplitude reconstruction over time for different methods, while (b) shows the PSNR values for phase reconstruction. The results for amplitude and phase are**

displayed on the right, with images (*3) to (*6) corresponding to different batch sizes respectively. "s" represents the sequential incremental gradient method and "GT" denotes the ground truth selected in the experiment.

We used an RTX 3090 graphics card with 24GB memory and a 4210R CPU for parallel operation. Fig. 4. plots the reconstruction accuracy versus time for different methods. When the batch size is one, it can be regarded as a stochastic gradient descent method. As described in Section 2.3, unlike previous stochastic gradient methods, a batch size of one ensures that each LR image is selected the same number of times. Compared to the original sequential update method, a batch size of one has better results. When the batch size is equal to the number of LR images, it can be regarded as global gradient descent, which has maximum parallelism, but convergence is more difficult and requires more iteration rounds. The global gradient descent method has a smoother curve due to moving towards the global optimum each time. Overall, the relationship between batch size and the number of iterations for our method can be simply described as: as batch size increases, algorithmic parallelism increases, single iteration time decreases, convergence becomes slower, and the overall number of iterations increases. This relationship makes the effect of batch size on the final convergence time negligible, as shown in Fig. 4. As a result, our method can be deployed on consumer graphics cards with varying memory capacities. The differences in batch size have almost no impact on the reconstruction time, which greatly reduces the computational cost. Overall, our method is about 10 times faster than the original method. This is expected to further promote the practical application of FPM in the WSI domain.

*3.2 Experiments on USAF resolution target.*

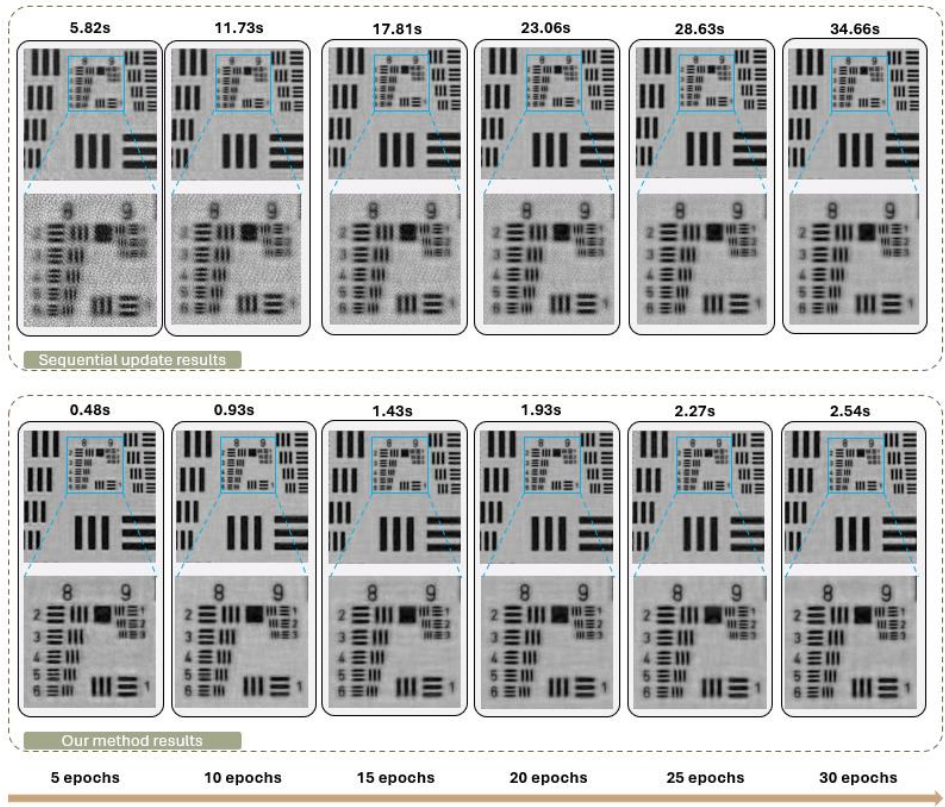

**Fig. 5. The results of USAF resolution target with different exposure time for each illumination angle.**

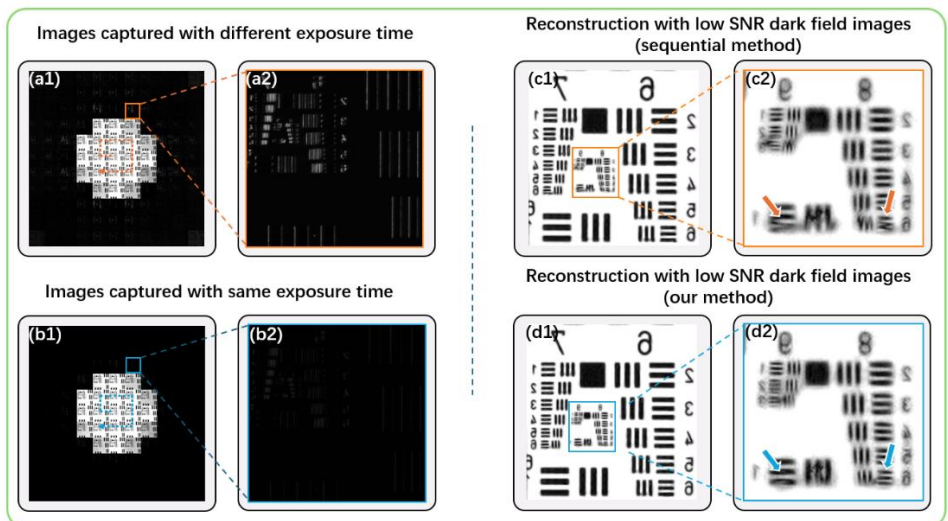

**Fig. 6. (a*) The captured LR images of USAF resolution target with different exposure time for each illumination angle. (b*) The captured LR images with the same exposure time. (c*) The reconstruction result of sequential method. (d*) The reconstruction result of our method.**

To verify the performance of our method in real experiments and to validate the theory in Section 2.3, we imaged the USAF resolution target using the same system parameters as in the simulation experiments. We set the batch size to 40 to reconstruct the LR images of $256 \times 256$ pixels to a HR image of $1024 \times 1024$ pixels. The results of our and the sequential methods with different optimization epochs are shown in Fig. 5. During the optimization process, our method tends to move towards the global optimum and demonstrates significantly stronger noise resistance. In other words, our method can achieve better reconstruction results with fewer iterations. In practice, our method can converge at a larger learning rate, usually 10 times larger than that of the sequential method. Sequential methods often struggle to converge at the same learning rate and may even diverge at the beginning of optimization, which supports the theory in Section 2.3. Our method successfully breaks the iterative loop mentioned therein and no longer relies solely on a reduced learning rate to converge to the stationary point. Overall, the increased parallelism during iteration and the reduced number of iterations together contribute to the speedup of our method.

To achieve clear reconstruction results, it is generally necessary to assign different exposure times to the low-resolution images captured under various illumination angles during the image acquisition process. For example, when using an sCMOS camera (PCO.edge 5.5, 6.5 μm pixel size) with a 0.13 NA 4x objective lens and maintaining the LED array at 97 mm from the sample, five different exposure times are typically employed. We encode the LEDs from positions 1 to 121 based on their illumination sequence: exposure times are set to 30 ms for bright-field images (LEDs 1-9), 150 ms for LEDs 10-25, 250 ms for LEDs 26-49, 350 ms for LEDs 50-81, and 450 ms for LEDs 82-121. To further demonstrate the robustness of our method, we reduced the dark-field exposure and used the same camera for image acquisition, keeping all other parameters consistent with the previous experiments. Unlike the usual approach, we set a uniform exposure time of 30 ms during the acquisition process. The results, shown in Figure 6, reveal that despite the significantly increased dark-field noise, our algorithm effectively reduced noise interference, confirming the analysis in Section 2.3 regarding the noise levels and sources associated with different optimization algorithms.

*3.3 Experiments on pathology sample*

To demonstrate the practical value of our method and its capability for multi-parameter joint correction in different regions of interest (ROIs), we observed H&E-stained colon cancer sections. We used an image sensor with a pixel size of 6.5 $\mu m$, a 10X objective lens with an NA of 0.3, and a 13 × 13 LED array for illumination at 52 mm from the sample. The computational resources remained the same as in the simulation experiments. We performed threshold noise reduction on the dark-field images by referring to the method of Zhang, Yan et al. [38]. To obtain the color images required for clinical diagnosis, we used light with wavelengths of 470 nm, 523 nm, and 623 nm for illumination, and then stitched the three reconstructed images at the channel level. We acquired 169 LR images of 2048 × 2048 pixels and performed multi-parameter joint correction reconstruction on different ROIs of 0.17 × 0.17 mm. The reconstruction results are shown in Fig. 7. Unlike previous algorithms that correct multiple parameters individually, our method controls the total imaging time for the monochrome ROI size to within 1 second. Overall, compared to many previous methods, our approach significantly improves temporal resolution and imaging robustness, demonstrating potential applications in clinical diagnosis, digital pathology, and life sciences.

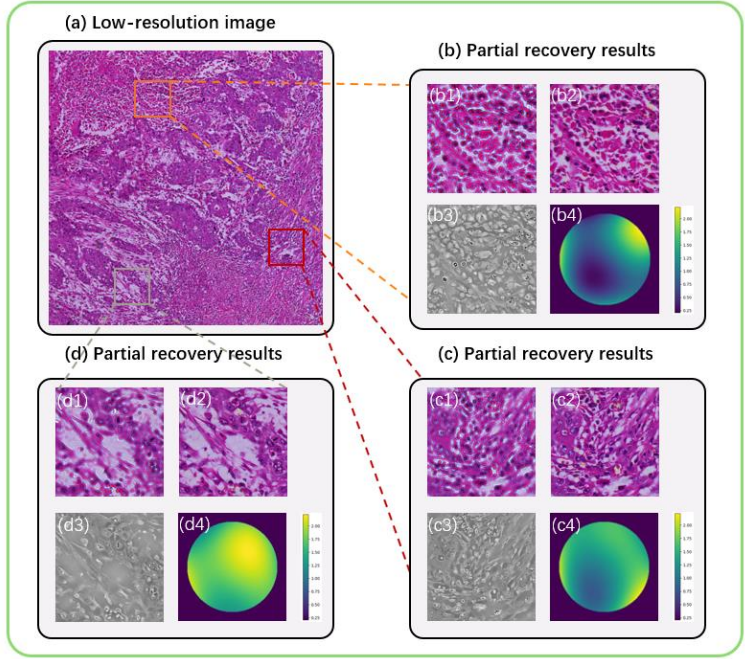

Fig. 7. Experimental results on H&E-stained pathological sections. (a) Captured low-resolution image illuminated by the center LED (b-d) The refocusing reconstruction results of different regions of interest.

## 4. Conclusion and discussion

In this study, we have introduced a novel fast and robust Fourier ptychographic microscopy (FPM) reconstruction method based on physical neural networks. Our method allows for simultaneous correction of multiple parameters, including defocus distance and pupil function, utilizing advanced optimizers from the field of deep learning. By revisiting the reconstruction process from an informatics perspective and adapting the model structure for batch-based optimization methods, we have significantly improved the temporal resolution and noise resistance of FPM without any hardware modifications. Our approach enhances GPU computational power utilization, effectively reducing the economic cost of image reconstruction. Experimental results demonstrate that our method achieves almost real-time digital refocusing reconstruction of a region of interest with a 1024 × 1024 pixels size on a relatively inexpensive consumer GPU. This advancement facilitates the widespread acceptance and application of FPM systems in clinical diagnostics, biomedical research, and digital pathology. Our work not only improves imaging robustness and speed but also assists researchers in conducting FP-related experimental and theoretical studies more efficiently. It can also be extended to other fields like Ptychographical Iterative Engine (PIE) and Coherent Diffraction Imaging (CDI).

Admittedly, our work still has some limitations. Although our method can directly reconstruct the acquired full-field images without the need for segmentation, there are some artifacts in the edge regions of the field of view. This is since the LED light source emits spherical waves, resulting in a significant tilt angle difference between the edge regions and the nearly parallel light incident at the center region, as well as variations in the pupil function across different regions. Previous methods typically required segmenting the full-field image for reconstruction, making it easier to correct these errors. To address this issue, we are actively exploring solutions, including but not limited to adding lenses between the light source and the sample to ensure the incident light is parallel. This will be one of the important directions for our future work.

**Caption List**

**Fig. 1 The figure shows the difference between our proposed method and previous sequential methods.**

**Fig. 2 Fast and robust Fourier ptychography multi-parameter neural network with batches update.**

**Fig. 3 The ground true of simulation.**

**Fig. 4** The simulation results.

**Fig. 5** The results of USAF resolution target with different exposure time for each illumination angle.

**Fig. 6** Experimental results at different exposure times.

**Fig. 7** Experimental results on H&E-stained pathological sections.